# A Multipath Energy-Aware On demand Source Routing Protocol for Mobile Ad-Hoc Networks

**S. Chettibi and M. Benmohamed**

Computer Sciences Dpt.,

University Mentouri of Constantine

Constantine, 25000 Algeria.

s.chettibi@yahoo.fr, ibnm@yahoo.fr

**Abstract**

*Energy consumption is the most challenging issue in routing protocol design for mobile ad-hoc networks (MANETs), since mobile nodes are battery powered. Furthermore, replacing or recharging batteries is often impossible in critical environments such as in military or rescue missions. In a MANET, the energy depletion of a node does not affect the node itself only, but the overall network lifetime. In this paper, we present multipath and energy-aware on demand source routing (MEA-DSR) protocol, which exploits route diversity and information about batteries-energy levels for balancing energy consumption between mobile nodes. Simulation results, have shown that MEA-DSR protocol is more energy efficient than DSR in almost mobility scenarios.*

**Keywords:** Routing Protocols, Mobile Ad-Hoc Networks, Energy Consumption.

## 1. Introduction

A mobile ad-hoc network (MANET) is a collection of mobile devices communicating in a multi-hop fashion, without relying on any fixed infrastructure or centralized authority. MANETs are very attractive for applications where establishment of a communication infrastructure is impossible (e.g. in battlefields), or even when there is just a need for a transient communication (e.g. in conferences).

There are three current areas of research in energy efficient routing in ad-hoc networks: 1) Power control [17,3], increases network capacity and reduces energy consumption by allowing nodes to determine the minimum transmit power level required to maintain network connectivity and forward traffic with least energy cost; 2) Maximum lifetime routing [13,14,7], selects paths that maximize network lifetime by balancing energy consumption across the nodes of the network; 3) Power save protocols [15,16,2], attack the problem of high idle state energy consumption by maximizing the amount of time nodes spend in the sleep state. Other works like [12,11], have combined both maximum lifetime routing and power control approaches.

In this paper, we present MEA-DSR a multipath and energy aware on demand source routing protocol for MANETs, based on DSR [5]. The remainder of this paper is organized as follows: in section 2, we give motivations to our work and we describe in some detail MEA-DSR protocol. We also evaluate MEA-DSR performances by simulations. Section 3 concludes the paper.

## 2. MEA-DSR protocol

### 2.1. Motivations

Routes failure is a norm rather than an exception in MANETs. Frequent route discoveries are needed to re-establish broken routes. Thus, a considerable global energy gain can be achieved by minimizing the frequency of route discoveries. In addition, although frequent topology changes permit some kind of load distribution by forcing nodes to discover new routes, but there is no guarantee that new routes be completely disjoint from the broken ones. Thus, some nodes still used, just because they have critical positions. Hence, an additional energy saving can be acquired by: 1) using node disjoint routes and 2) by taking into account battery levels of nodes when making routing decisions.

## 2.2. MEA-DSR protocol

MEA-DSR protocol is based on DSR [5]. In fact, DSR is a multipath routing protocol. However, in DSR multiple routes are stored in a trivial manner with no constraint on number or quality. MEA-DSR, limits the number of routes that a destination node provides to a source node to two. It was shown in [9] that the performance advantage from using more than one or two alternate routes is minimal.

The choice of the primary route in MEA-DSR is conditioned by two factors: 1) the residual energy of nodes belonging to the route; 2) the total transmission power required to transmit data on this route. If we consider that nodes transmit all with their maximal transmission power (adjusting transmission power feature is not supported by all network interface cards), the later factor is equivalent to that of hops number in the route. Concerning the choice of the second route, the disjunction ratio from the primary route comes in first order. If several routes present the same disjunction ratio, one will be chosen via the same criterion as for the primary route.

Instead of splitting traffic on several routes, only one route is used in MEA-DSR, during a communication session until its breakage. This permits to avoid problems of: route coupling [10], congestion of common nodes and out of order arrival of data packets to their destinations.

## 2.3. Control packets and data structures used in MEA-DSR protocol

Mobile nodes using MEA-DSR, exchange three types of control packets: route requests (RREQs), route replies (RREPs) and route errors (RRERs). The same format of RRER and RREP defined in [6] for DSR is used in MEA-DSR, whereas RREQ format has been slightly modified. MEA-DSR utilises three data structures: routes cache, route requests table and routes table. The same structure defined in [6] for routes cache is reutilised in MEA-DSR. Route requests table format has been enriched by additional fields; routes table is a new data structure specific to MEA-DSR.

### 2.3.1. RREQ Packet

A field called « *min_bat_lev*» has been added to RREQ packets. It takes as value the minimum of residual energies of nodes traversed by the RREQ packet.

### 2.3.2. RREQs table

The format defined in [6] has been augmented by the following fields: 1) «*nb_hops*» indicates the number of nodes traversed by the RREQ; 2) « *last_node*» maintains the identifier of the neighbor who transmitted the RREQ.

### 2.3.3. Routes table

This structure is utilised to store every candidate route in destination nodes, indexed by source node identifier. Every entry in the routes table contains the following fields:
- *Src: maintains the identifier of source node who initiated the route discovery procedure.*
- *Seq: maintains the RREQ sequence number.*
- *Route: contains the nodes sequence traversed by RREQ packet.*
- *Min_bat_lev: keeps the minimal residual energy of nodes traversed by RREQ packet.*
- *Arrival_time: keeps the arrival time of RREQ packet at the destination node.*

Content of the first four fields is directly extracted from arriving RREQ packets.

## 2.4. Description of MEA-DSR protocol operation

MEA-DSR is composed of three phases: 1) route discovery; 2) route selection and ; 3) route maintenance phase.

### 2.4.1. Route discovery

If a source node needs a route toward a destination and no one is stored in its cache, it broadcasts a RREQ to all its neighbors. In MEA-DSR, only destination nodes can respond to a RREQ packet, because it will be difficult to control route disjunction if intermediate nodes reply directly from their caches as it is the case in DSR.

In order to avoid overlapped route problem [8], intermediate nodes do not drop every duplicate RREQs and forward duplicate packets coming on a different link than the link from which the first RREQ is received, whose hop count is not larger than that of the first received RREQ. However, forwarding all duplicates satisfying this criterion generates a very high overhead. Thus, we have limited the number of copies to be forwarded to one.

When an intermediate node situated in the neighborhood of source node, receives the RREQ packet it includes its residual energy value in *«min_bat_lev»* field. Otherwise, any intermediate node compares its residual energy to the value of *«min_bat_lev»* field; if it is lower, it changes the value of *«min_bat_lev»* by its proper value. After the end of above procedure, the intermediate node appends its identifier to the RREQ packet and rebroadcasts it to its neighbors. This process will be continued until that the RREQ packet arrives to destination.

### 2.4.2. Route selection

After reception of the first RREQ packet, the destination node waits for a certain period of time "Wait_time" before starting route selection procedure. When this period of time expires, destination node selects as a primary route *'route$_i$'*, satisfying the following condition:

$$\frac{\min\_bat\_lev_i}{route\_length_i} = \max_{j=1,n}\left(\frac{\min\_bat\_lev_j}{route\_length_j}\right) \quad (1)$$

Where, *n* is the number of candidate routes stored in the routes table.

After the selection of the primary route, destination node sends immediately a route reply to the source node. The alternative route must be maximally node disjoint than the primary route. If there exist several routes with a same disjunction ratio, one that satisfies equation (1) will be chosen and included in a RREP to be send to the source node.

### 2.4.3. Route maintenance

In MEA-DSR, if an intermediate node detects a link failure, it transmits a RERR message to the upstream direction of the route. Every node receiving the RERR message, removes every entry in its route cache that uses the broken link, and forwards RERR message to the next node toward the source node. If the source node has no valid route in its cache, then it reinitiates a new route discovery phase.

### 2.5. Simulation

### 2.5.1. Simulation environment

We have used NS-2 simulator [18], for MEA-DSR performance evaluation. The studied network is a collection of 50 nodes deployed on square area of 1000mx1000m. Each node has a transmission range of 250 m. The medium access control (MAC) protocol was based on IEEE 802.11 with 2 Megabits per second raw capacity. For radio propagation model, a two-ray ground reflection model was used. In all simulations, we have utilized the RWP (Random waypoint) mobility model [1]. Each node moves with a maximum speed randomly chosen from the interval [5 m/s, 10 m/s]. The duration of every simulation was 600 seconds, executed with different mobility scenarios characterised by different pause times.

Communication between nodes was modelled by CBR (Constant Bit Rate) traffic over UDP. A source generates packets of 512 bytes with a rate of four packets per second. A total of 10 connections were generated. They start at a time randomly chosen from the interval [0s, 120s] and still active until the end of simulation.

It was shown in [7], that no real optimisation can be achieved in the presence of overhearing. For this reason, we have only considered energy consumed in transmission and reception modes. As value we have utilised those obtained trough experiments in previous works [4] (1.4 W for transmission mode and 1 W for reception mode).

MEA-DSR, was compared with DSR. Furthermore, it was simulated for different value of "Wait_time".

### 2.5.2. Performance metrics

In simulations, we are interested in the following performance metrics
- Packet delivery fraction (PDF)- the ratio of data packets well received by destination nodes to those generated by source nodes.
- Average end to end delay (AD)- the average time that takes a data packet from the source node to the destination node.

- Normalized routing overhead (NRO)- the ratio of the number of routing protocol control packets (*RREQs, RREPs, and RRERs*) transmitted to the number of data packets received.
- Consumed energy per packet (CEP) -the ratio of global consumed energy to the number of data packets received
- Standard deviation of consumed energy per node (SDCEN)- square root of the average of the squares of the difference between the energy consumed at each node and the average energy consumed per node.

### 2.5.3. Simulation results

We use the notation 'MEA-DSR-nWT' to indicate MEA-DSR protocol with any value for 'Wait_Time'.

#### 2.5.3.1. Packet delivery fraction

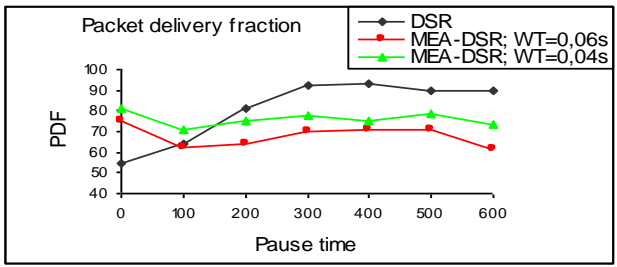

**Figure 1.** Packet delivery fraction vs. pause time

Under high mobility scenarios, MEA-DSR-nWT presents a higher packet delivery fraction than DSR. The reason is that intermediate nodes in DSR, are authorized to answer from their caches. However, in such mobility conditions stored routes are more likely to be stale. Thus, data packets forwarded on those routes will be dropped, as soon as they reach broken links, since also salvaging mechanism becomes less efficient. In addition, data packets are more likely to expire because of the additional latency introduced by frequent retransmissions and repetitive salvaging attempts. For lower mobility scenarios, the packet delivery fraction of DSR increases to surpass MEA-DSR-nWT one, because routes tend to be more robust and both responding from cache and salvaging mechanisms become more efficient. In MEA-DSR-nWT, intermediate nodes are not authorized to use their caches to salvage data packets. Thus, data packets probability to be dropped is greater than it is in DSR.

MEA-DSR-nWT shows its maximum packet delivery fraction in the case of constant motion (pause time=0), this can be explained by routes variety imposed by mobility. This variety decreases the risk of queue congestion and thus of data packets expiration. For moderate to low mobility scenarios, MEA-DSR-nWT presents approximately the same behaviour, because it uses relatively long stable routes. Long routes are more stable than short ones, since the average ratio of physical link distance to the transmission range of a hop in a short route is greater than it is in a long route. Thus, shortest routes tend to break earlier.

For increased values of WT, MEA-DSR shows a lower packet delivery fraction. The reason is that MEA-DSR tends to use longer routes. Packets traversing longer routes spend more time when forwarded from one interface queue to another, which can lead to their expiration.

#### 2.5.3.2. Average end-to-end delay (AD)

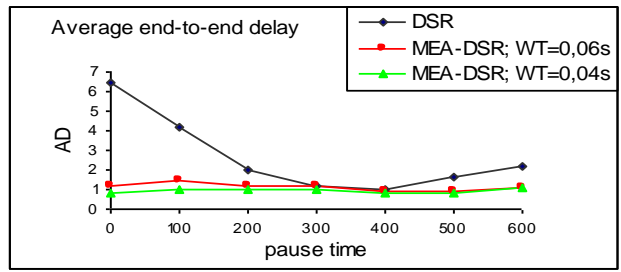

**Figure 2.** Average end-to-end delay vs pause time

Average end to end delay in DSR and under all mobility scenarios, is always higher than it is in MEA-DSR-nWT. In high mobility conditions, data packets in DSR spend more time in interface queues due to frequent retransmissions and repetitive salvaging attempts. In low mobility scenarios, intermediate nodes in DSR respond from their caches. Their replies tend to contain long routes since they make concatenations. Thus, data packets are queuing several times before reaching their destinations.

It is very clear on fig. 2, that average end-to-end delay in MEA-DSR-nWT is independent of mobility. This is because MEA-DSR-nWT tends to use relatively stable routes of approximately same length. For growing values of WT, the average end to end delay in MEA-DSR has slightly increased. This was expected, because a destination node takes more time to reply.

### 2.5.3.3. Normalized routing overhead

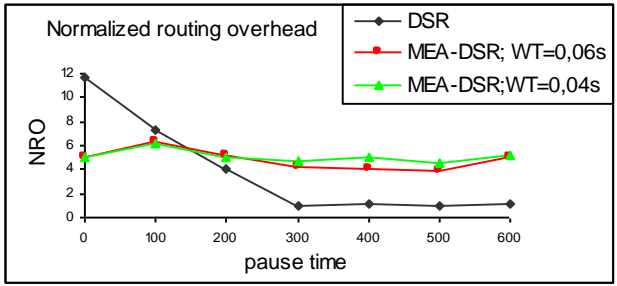

**Figure 3.** Normalized routing overhead vs pause time

Under high mobility scenarios, DSR generates more routing overhead than MEA-DSR-nWT, because DSR reinitiates route discoveries repeatedly and consequently generates more RREQs. In lower mobility scenarios, routes become more stable. Thus, the need to reinitiate new route discoveries is diminished for both protocols. However, MEA-DSR-nWT stills generate high overhead. This is because MEA-DSR-nWT permits intermediate nodes to propagate duplicates of RREQs, whereas in DSR intermediate nodes drop every duplicate RREQ packet.

For growing values of WT, routing overhead in MEA-DSR decreases because greater is WT, greater is the number of RREQs received at destination nodes. Hence, it is more likely to discover alternate routes with higher disjunction ratio, diminishing the risk of simultaneous failure of both primaries and alternates routes.

### 2.5.3.4. Consumed energy per packet

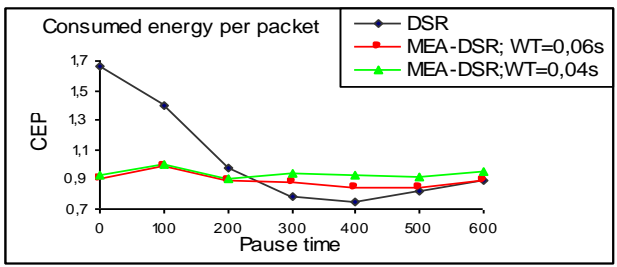

**Figure 4.** Consumed energy per packet vs pause time

Consumed energy per packet gives an idea about the global energy consumption in the network; it is clear that is proportional to the generated routing overhead. For high mobility scenarios, DSR generates more overhead than MEA-DSR-nWT thus it consumes more energy. For lower mobility scenarios, although DSR generates less overhead but it does not present an important improvement in energy consumption because it tends to use longer routes (total transmission power of a packet stills high).

### 2.5.3.5 Standard deviation of consumed energy per node

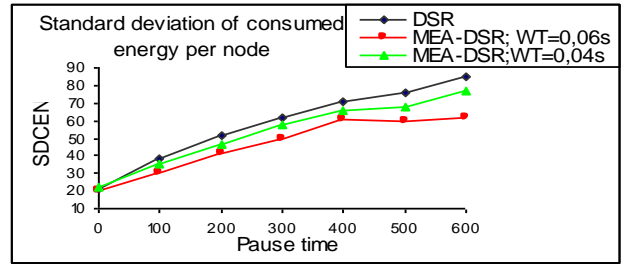

**Figure 5.** Standard deviation of consumed energy vs pause time

Standard energy deviation of consumed energy per node gives an idea about the fairness of node usage by routing protocols, and so about routing protocol tendency to maximize network lifetime. Under all mobility scenarios, standard energy deviation of consumed energy per node in MEA-DSR-nWT is always lower than that of DSR, which confirms the efficiency of load distribution policy adopted in MEA-DSR-nWT (choice of routes having nodes rich in energy and the use of maximally node disjoint routes). For both protocols, network stability provokes standard deviation of consumed energy increase. This was expected, since routes still in use in a communication session while they are valid.

The higher is WT, the higher is the efficiency of load distribution policy of MEA-DSR since destination nodes receive more RREQs. Therefore, it is more likely to discover routes with higher minimal residual energy and to discover alternate routes with higher disjunction ratio.

## 3. Conclusions

Node-disjoint routes are exploited in MEA-DSR to 1) achieve a global energy gain by minimizing frequent route discoveries; and to 2) balance energy consumption of mobile nodes. The choice of the primary route in MEA-DSR is dictated by minimal residual node energy to route length ratio, whereas disjunction ratio from primary route comes in first order in alternate route choice. Simulation results have shown that under high mobility scenarios, MEA-DSR is better than DSR that is in network operation performances or in energy efficiency. For lower mobility scenarios, MEA-DSR stills more energy efficient than DSR, whereas it presented lower packet delivery fraction and higher routing overhead. It was also shown that the higher is WT (wait_time) the higher is energy efficiency and the lower is packet delivery fraction and vice-versa.

## Web links